\documentclass[final]{agujournal2019}
\usepackage{url} 
\usepackage{lineno}
\usepackage[inline]{trackchanges} 
\usepackage{soul}
\draftfalse

\journalname{AGU Space Weather}

\begin{document}

%
%


\title{The Effect of Field Line Tracing Height on Wang-Sheeley-Arge Solar Wind Predictions at Earth}

%
%




\authors{Satabdwa Majumdar\affil{1}, Martin A. Reiss\affil{2,3}, Pete Riley\affil{4}, Charles N. Arge\affil{5}}


\affiliation{1}{Austrian Space Weather Office, GeoSphere Austria, Graz, 8020, Austria}
\affiliation{2}{Community Coordinated Modeling Center, NASA Goddard Space Flight Center, 8800 Greenbelt Rd., Greenbelt, MD 20771, USA}
\affiliation{3}{USRA}
\affiliation{4}{Predictive Science Inc., San Diego, California, USA}
\affiliation{5}{NASA Goddard Space Flight Center, 8800 Greenbelt Rd., Greenbelt, MD 20771, USA}





\correspondingauthor{Satabdwa Majumdar}{satabdwa.majumdar@geosphere.at}




\begin{keypoints}
\item Raising the magnetic field line tracing height above the photosphere improves the prediction of solar wind high-speed enhancements at L1.
\item It reduces small-scale fragmentation and produces smoother, more spatially coherent modeled open/closed field boundaries.
\item Multi-metric validation reveals performance improvements that are not captured by standard point-to-point evaluations alone.
\end{keypoints}

%
%

%
%


\begin{abstract}
A key component of the widely used Wang-Sheeley-Arge model is the tracing of magnetic field lines in the solar corona, as that largely determines the modeled open/closed field regions and in turn the predicted large-scale solar wind structure. However, small closed field regions interspersed among the open field lines inside coronal holes (CHs) often hinder determination of the large-scale open–closed boundary. Since most of this small closed flux inside CHs is confined to low heights, we demonstrate that by slightly raising the height from which magnetic field lines are traced from the photosphere (R$_{su}$), we can improve the prediction of solar wind high-speed enhancements (HSEs). We find a higher R$_{su}$ produces more accurate estimates of the HSE start time, duration, and peak speed, and enables the detection of HSEs that were missed with the standard R$_{su}$. These improvements are primarily associated with changes in the great-circle angular distance, while the flux tube expansion factors remain largely unchanged. A higher R$_{su}$ produces smoother open-field boundaries and reduces small-scale fragmentation within open-field regions. These morphological changes lead to more stable magnetic footpoint mappings and consequently more robust GCD calculations. To assess the statistical relevance of these findings, we extend the analysis to 53 Carrington rotations and perform a multi-metric evaluation of model performance. The results indicate that higher R$_{su}$ improves HSE predictions while reducing false alarms. We thus find that the choice of R$_{su}$ represents an important and previously underexplored parameter that can help enhance our ambient solar wind modeling and forecasting capabilities.
\end{abstract}

\section*{Plain Language Summary}
Accurately predicting the solar wind is essential for forecasting space weather near Earth. A commonly used model (the Wang-Sheeley-Arge model) traces magnetic field lines in the Sun’s atmosphere to identify regions (called coronal holes) that produce fast solar wind. However, small magnetic structures within these regions can fragment coronal holes, making it difficult to accurately determine where fast wind originates. In this study, we investigate a simple method for improving solar wind predictions by slightly increasing the height above the Sun’s surface from which magnetic field lines are traced. We find this reduces the effect of small-scale fragmentation and produces smoother open-field boundaries. As a result, the model more accurately predicts when high-speed solar wind streams arrive, their duration, and how fast they are. It also improves detection of previously missed events and reduces false alarms. Using multiple evaluation methods across many solar rotations, we show that this small change can lead to more reliable solar wind forecasts in the future.

%
%

%


%
%
%
%

\section{Introduction} \label{intro}

Accurate ambient solar wind modeling is essential for a thorough understanding of heliospheric dynamics and for providing reliable space weather forecasts at Earth and throughout the inner heliosphere \cite<>[]{Verscharen2019LRSP}. Coronal Mass Ejections, which are major drivers of severe space weather, propagate through this ambient wind and experience a drag force that can substantially affect their morphology and kinematics \cite{Gopalswamy2000GeoRL,Vrsnak2013SoPh,Majumdar2020ApJ,Majumdar2021ApJ}. Solar energetic particles, which are also major sources of ground level enhancements \cite<>[]{Reames2023FrASS}, propagate along pathways defined by the magnetic field embedded in the solar wind. In addition, the solar wind exhibits substantial variability in its plasma properties, including its speed, density, temperature, composition, and ionic charge states, reflecting the diverse physical processes involved, which further motivates the need for a comprehensive understanding of its origin and evolution \cite<>[]{Cranmer2017SS}.

To achieve a good understanding of the solar wind, scientists often employ a series of coupled modeling units to model the whole heliospheric solar wind. Such a modeling framework often consists of a coronal domain, a heliospheric domain, and sometimes the two joined together by an interface region \cite<see, for instance, >[]{Reiss2023AdSpR}. The coronal domain commonly uses a synoptic map of the photospheric magnetic field as input. Since routine measurements of the coronal magnetic field are still limited, the coronal domain relies on extrapolation techniques to reproduce the large-scale coronal magnetic field \cite<see >[]{Wiegelmann2017SSRv}. One of the most widely-applied extrapolation techniques in the coronal domain is  the Potential Field Source Surface (PFSS) model that assumes a current-free potential field configuration of the corona up to a certain height called the source surface height \cite<>[]{Altschuler1969SoPh,Schatten1969SoPh}. The interface domain defines the outer boundary conditions of the coronal domain and the inner boundary conditions of the heliospheric domain. In this interface, empirical relationships are used to specify the solar wind speed at this boundary. Notable examples include the Wang-Sheeley model \cite<WS; >[]{Wang1990ApJ}, the Distance from Coronal Hole Boundary model \cite<DCHB; >[]{Riley2001JGR} and the Wang-Sheeley-Arge model \cite<WSA; >[]{Arge2004JASTP}. The WS model is based on an inverse relationship between the flux-tube expansion factor ($f_{exp}$) of the open magnetic field lines and the solar wind speed measured at Earth, while the DCHB model is based on the relationship between the great circle angular distance (GCD) to the nearest open-closed boundary and the solar wind speed at the solar surface. The WSA model stitches together the above two relationships and provides a solar wind speed that relies on both the EFs and the GCDs. The resulting solar wind speed at the outer boundary of the interface domain then serves as an inner boundary condition for heliospheric propagation models, which transport the solar wind from the near-Sun environment into the heliosphere and to 1 AU. In this study, we employ the Heliospheric Upwind Extrapolation model \cite<HUX; >[]{Riley2011SoPh} to perform this propagation. The HUX model is a simple 1D upwind model that simplifies the fluid momentum equation by neglecting the pressure gradient and gravitational terms, and allows the solar wind to evolve and interact in a self-consistent manner. Recent developments have further extended the capabilities of this model through the introduction of the Tunable Heliospheric Upwind Extrapolation model \cite<THUX; >[]{Reiss2020ApJ} and the time-dependent Heliospheric Upwind Extrapolation model \cite<HUXt; >[]{Owens2020SoPh,Barnard2022FrP}. Such a PFSS–WSA–HUX modeling framework has been widely adopted in recent years for simulating the solar wind \cite<>[]{Reiss2019ApJS,Reiss2020ApJ,Mayank2022ApJS,Issan2023SpWea,Majumdar2025ApJ} not only for its computational efficiency but also for its reasonable accuracy compared to other state-of-the-art modeling frameworks. Furthermore, \citeA{Heinemann2025ApJ} demonstrated, using HUXt, that the model is comparatively less sensitive to variations in the solar wind inner boundary conditions at 0.1 AU than other heliospheric models. For a review on the different heliospheric solar wind models, see \citeA{MacNeice2018SpWea}.

In the fluid momentum equation, the dynamic pressure term ($\sim \rho v^2$) plays a dominant role in shaping the structure and evolution of the solar wind. Consequently, inaccuracies in the near-Sun solar wind speed map ($v$), produced by the coupled coronal and interface domains, can propagate outward and become amplified, leading to significant degradation in the accuracy of heliospheric solutions \cite{Gonzi2021SpWea}. This underscores the importance of identifying and quantifying the sources of errors and uncertainties in the modeled solar wind speed within the coronal and interface modeling stages, as improvements in these stages directly translate into more reliable solar wind simulations and predictions. In the WSA framework, accurately modeling the solar wind at L1 critically depends on a reliable determination of the large-scale open/closed magnetic field regions, as these directly govern the large-scale solar wind structure near the Sun and, consequently, throughout the heliosphere. These modeled open field regions can be affected either by the choice of the input magnetic map \cite{Riley2012SoPh,Li2021JGRA,Demidov2023SoPh,wang2024ApJ,Kumar2025JSWSC} or by the choice of the coronal model (e.g. PFSS, non linear force free, MHD) and the coronal model settings. In the case of the coronal model settings, lowering the source surface height (Rss) from its standard value of 2.5 R$_{\odot}$ has been shown to better reproduce observed large-scale open-field boundaries in some cases \cite{Lee2011SoPh,Nikolic2019SpWea,Asvestari2019JGRA,Badman2022ApJ}, or can better match modeled open field regions from a magnetohydrodynamic model \cite{Huang2024ApJ} although the statistical significance of the overall agreement is still debatable. However, it should be noted that such agreement is often achieved at the cost of opening up spurious open field regions that do not have any associated coronal holes in extreme-ultraviolet observations. Further, in the coronal domain, to improve WSA predictions, studies have focused on optimizing the empirical coefficients in the WSA empirical relation \cite{McGregor2008JGRA,Riley2015SpWea,Reiss2020ApJ,Mayank2025ApJ} or the model grid resolution \cite{Riley2013JGRA,Majumdar2025ApJ}. Building on this, recent studies have also demonstrated that an optimized Rss can further improve solar wind predictions at L1 \cite{Meadors2020SpWea,Majumdar2025ApJ,Kumar2025JSWSC}. In particular, \citeA{Majumdar2025ApJ} proposed generating an ensemble of solar wind solutions corresponding to different Rss values, wherein the resulting open field regions are constrained using coronal hole observations in extreme-ultraviolet, keeping in mind the inherent uncertainties in the observed coronal hole boundaries as summarized in \citeA{Reiss2024ApJS}. 

\begin{figure*}[h]
\centering
\noindent\includegraphics[width=0.7\textwidth]{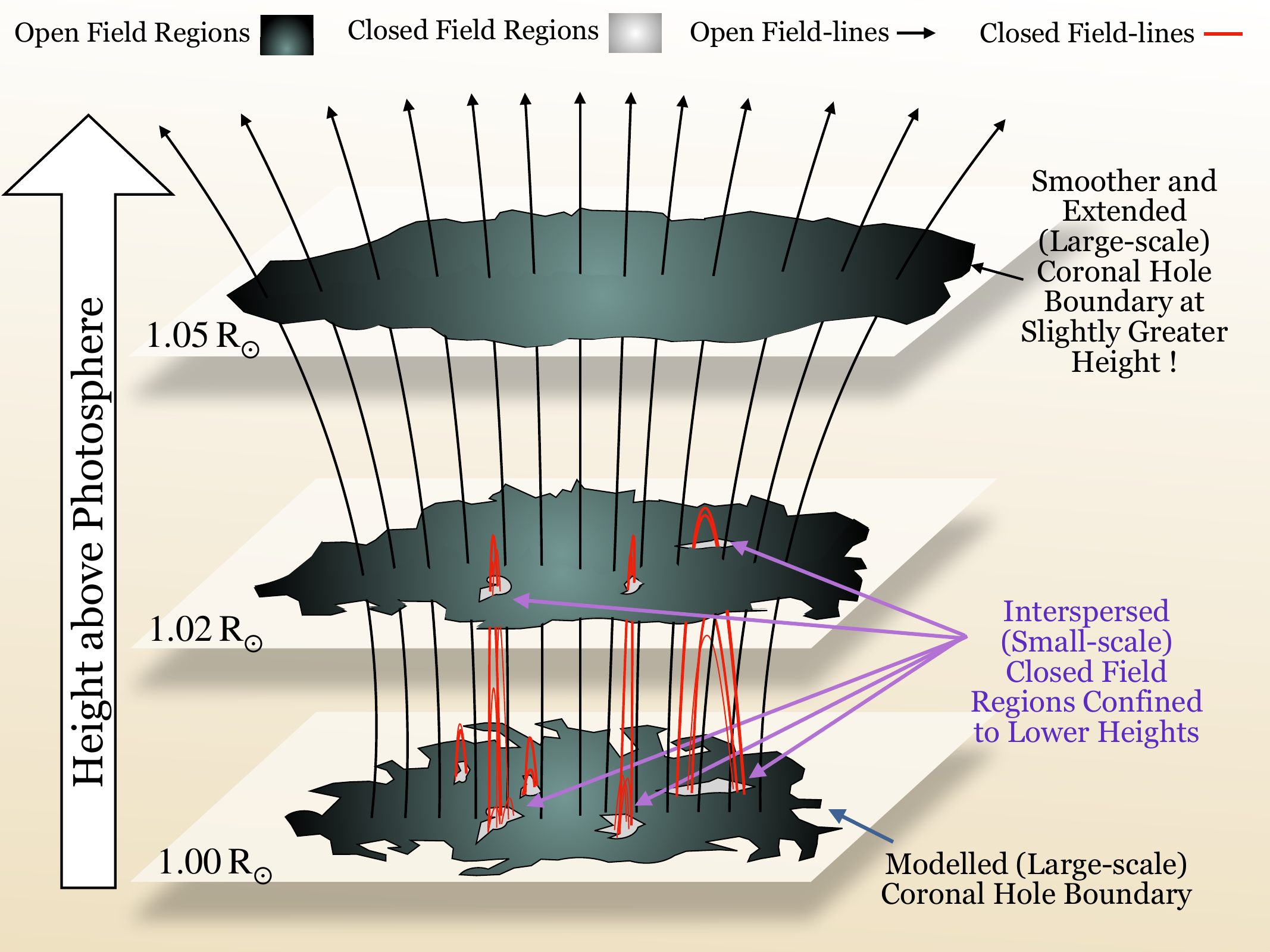}
\caption{A simplified representation (not to scale) of the idea of tracing magnetic field-lines from two slightly different greater heights (1.02 and 1.05 R$_{\odot}$) than the photosphere. The mentioned heights are chosen as an example to demonstrate the idea of the study. The dark green regions indicate the modeled open-field regions and the grey patches inside denote the small-scale closed-field regions that are mainly confined to lower heights inside the open-field regions. The black and red lines indicate open and closed field-lines respectively.}
\label{rsu_schematic}
\end{figure*}

Now, while the choice of the source surface height affects the predicted solar wind conditions at L1, it also introduces additional uncertainties in determining the origins of fast solar wind streams \cite{Koukras2025A&A}. In addition, accurately characterizing open field regions (which in turn determines the large-scale solar wind structure) remains challenging due to the complex multi-scale magnetic structure of the low corona. \citeA{Wiegelmann2004SoPh} showed that coronal holes contain an abundance of very low-lying closed loops. The presence of these small-scale closed-field structures intermittently embedded within the open field regions inside coronal holes (see Figure~\ref{rsu_schematic}) often makes it difficult for models like the PFSS model to identify the large-scale open field boundaries \cite{Heinemann2026ApJ}, leading to fragmented modeled open field regions. This ambiguity is particularly crucial for the WSA model for determining the GCD parameter, which is highly sensitive to the precise location of these boundaries. However, since most of these small-scale closed flux are confined to very low heights, a more reliable estimate of the large-scale open field boundary location can be achieved by evaluating the open flux distribution at slightly higher heights, where the influence of such small-scale closed structures is diminished and the magnetic topology appears more coherent. A very simplified schematic (not to scale) demonstrating this physical picture is shown in Figure~\ref{rsu_schematic}. Thus, it becomes imperative to investigate whether modeling the large-scale open field boundaries at a slightly greater height would improve the model's accuracy in predicting high-speed enhancements (HSEs) at L1. We broadly define these HSEs as intervals of elevated solar wind speed, irrespective of whether they satisfy the full set of plasma signatures typically used to identify classical high-speed streams (HSSs, see \citeA{Jian2019SoPh,Grandin2019JGRA}). We consider HSE as any region
of solar wind, 2 days or longer, in which the net solar wind
speed increase is 100 km/s or more \cite<see >[]{Owens2005JGRA}. This distinction is motivated by the fact that our modeling framework primarily predicts the solar wind speed profile and does not explicitly reproduce the additional plasma and magnetic signatures employed in HSS identification. Motivated by the above idea, in this paper, we thus investigate the effect of modeling the large-scale open field regions by tracing the magnetic field lines from a height slightly higher than the photosphere, and how this change affects the solar wind predictions at L1. In Section~\ref{data}, we describe the data and methodology of this study; this is followed by the results in Section~\ref{results}. We discuss the relevance of the results in Section~\ref{discussions}, and finally, we summarize the main conclusions from this study in Section~\ref{conclusions}.

\section{Data and Methods} \label{data}
As input to our modeling framework, we use full-disk GONG/NSO photospheric magnetic field maps, which are measured in Gauss, and are given in a $sin(\theta)-\phi$ grid, with 180 $\times$ 360 grid points, where $\theta\,\in$ ($-90^{\circ}, 90^{\circ}$)  is the heliographic latitude and $\phi\,\in$ ($0^{\circ}, 360^{\circ}$)  is the Carrington longitude, respectively. These magnetic maps are available at their online platform (https://gong.nso.edu/), either as near-real-time maps or full Carrington Rotation (CR) maps. In this study (similar to our previous study \citeA{Majumdar2025ApJ}), we use the CR maps. It should be noted here that the choice of the input magnetic map can also influence the modeled open-field regions, but such an investigation is outside the scope of the present study. To validate the model results with in-situ measurements, hourly-averaged observed solar wind speed data from the OMNIWeb database (https://omniweb.gsfc.nasa.gov/ow.html)  have been used \cite{Papitashvili2014}.

We use a similar three-component modeling framework as in our previous study \cite<>[]{Majumdar2025ApJ}: the coronal domain, a model interface and the heliospheric domain. In the coronal domain, we use the PFSS model \cite{Altschuler1969SoPh}, which in this study is the Python module \textit{pfsspy} \cite{Stansby2020}. This module uses a finite difference method \cite<based on >[]{van2000ApJ} to calculate the magnetic vector potential in a uniformly spaced grid in ln(r), $\mathrm{cos(\theta),\, \phi}$ in the spherical coordinates (r, $\mathrm{\theta},\,\phi$). We use 180 and 360 model grid points in the latitude and longitude directions respectively and 35 grid points in log(r) in the radial direction. The magnetic field lines between the inner boundary (with radius R$_{su}$) and the outer boundary (the source surface (with radius Rss)) of the coronal model are traced in two ways. We trace the field lines from the inner boundary to the outer boundary, where both open and closed field lines are traced to obtain the modeled open field regions. We also trace the open field lines starting from the outer boundary to the inner boundary, where the field lines end up inside the boundaries of the modeled open field regions. Now, for the case of tracing the field lines, instead of using the standard photospheric height (1 R$_{\odot}$) as the inner boundary, we use 6 different initial tracing heights (1.00, 1.01, 1.02, 1.03, 1.04, 1.05 R$_{\odot}$). Increasing the tracing height for the inner boundary in this manner reduces the influence of low-lying closed field structures embedded within open-field regions, thereby mitigating small-scale fragmentation both within the open-field region interiors and along their boundaries. The large-scale coronal magnetic structure thus found from the PFSS model is then fed to the interface region which consists of the Wang-Sheeley-Arge (WSA) empirical relation \cite{Arge2003AIPC,Arge2004JASTP} that generates the solar wind conditions near the Sun at the outer boundary of the PFSS model (at Rss). This empirical relation unifies two important aspects of the coronal magnetic field to derive the solar wind speed, which are the magnetic flux-tube expansion factor ($f_{exp}$) \cite{Wang1990ApJ}, and the great circle angular distance (GCD) of the open field lines footpoints from the nearest open-closed boundary. This GCD is also the essential component of the Distance from Coronal Hole Boundary model \cite<DCHB; >[]{Riley2001JGR}. The WSA relation unifies these two magnetic features to provide the solar wind speed through the following relation,

\begin{equation}\label{eqn-v_wsa}
    v_{wsa}(f_{exp},d) = c_1 + \frac{c_2}{(1+f_p)^{c_3}} \biggl\{c_4 - c_5exp \biggl[-\biggl(\frac{d}{c_6}\biggr)^{c_7}\biggr] \biggr\}^{c_8}
\end{equation}

where $f_{exp}$ and $d$ are the expansion factor and GCD respectively and $c_1$ to $c_8$ are model coefficients, which are taken from \citeA{Reiss2019ApJS,Reiss2020ApJ,Majumdar2025ApJ} as $c_1 = 285$, $c_2 = 625$, $c_3 = 2/9$, $c_4 = 1$, $c_5 = 0.8$, $c_7 = 2$, $c_8 = 3$ and $c_6$ is extracted as the median value of the GCD profile taken at the sub-Earth points at the source surface \cite<for more details, see >[]{Majumdar2025ApJ}. The near-Sun solar wind speed is then computed from Equation~\ref{eqn-v_wsa}, which is then fed into the heliospheric modeling domain. For the heliospheric domain, we use the Heliospheric Upwind eXtrapolation (HUX) model \cite{Riley2011SoPh}, which uses a 1D upwind extrapolation technique that simplifies the fluid momentum equation by neglecting the pressure gradient, gravity, and the magnetic field terms. The computational efficiency of HUX \cite{Riley2011SoPh} makes it particularly suitable for this study, which requires multiple model runs across different CRs (318 runs in total for 53 CRs) to systematically investigate the effect of increasing R$_{su}$ on solar wind predictions. Such a study would be considerably more computationally demanding with a fully physics-based heliospheric model. Besides its computational efficiency, \citeA{Owens2020SoPh} demonstrated close agreement between steady-state solar wind speed solutions obtained with the physics-based heliospheric version of the Mag-
netohydrodynamics Algorithm outside a Sphere model \cite<HelioMAS; >[]{Linker1999JGR,Riley2001JGR} and HUXt, with a statistical disagreement of less than 6.4$\%$ over a period exceeding 40 years. This further supports the suitability of the HUX model for the current analysis by offering reasonable accuracy at low computational costs. In the final step, we compare our modeled results with in-situ solar wind data from the hourly-averaged OMNI database. A Schatten Current Sheet (SCS) model \cite{Schatten1971CosEl} is often used in conjunction with the PFSS model for modeling the large-scale coronal magnetic field. However, conflicting boundary conditions at the interface between the two models often lead to tangential discontinuities in field lines or "kinks". In this context, \citeA{Wang1988JGR} showed that such kinks have minimal effect on the overall coronal solution, while \citeA{McGregor2008JGRA} later demonstrated that they do not affect the predicted solar wind speeds most of the time. Since the primary objective of this study is to investigate the effect of R$_{su}$ while keeping the remaining components and parameters of the modeling framework unchanged, we did not include an SCS model. Nevertheless, it would be worthwhile to investigate whether the effects of R$_{su}$ identified in this study remain consistent within a modeling framework that includes an intermediate SCS model. Such an investigation could be incorporated into the planned tests of the future real-time implementation discussed in Section 4.4. As previously discussed by \citeA{Majumdar2025ApJ}, we consider some key characteristics of an HSE. These are the starting time of the HSE ($t_{start}$), the duration of the HSE ($t_{dur}$), which we define as the interval between the start and end time of the HSE, and the maximum speed attained by the HSE ($v_{peak}$). Thus, a good prediction is one that minimizes the difference between modeled and observed values in these quantities. Throughout this study, we use a model grid resolution of 1$^{\circ}$ in both latitude and longitude as a higher resolution helps in minimizing errors in t$_{start}$ \cite<see >[]{Majumdar2025ApJ}. In what follows, we show for the first time how changing the initial tracing height of the magnetic field lines (R$_{su}$) affects solar wind modeling results at L1. 

\begin{figure*}[h]
\includegraphics[width=0.9\textwidth]{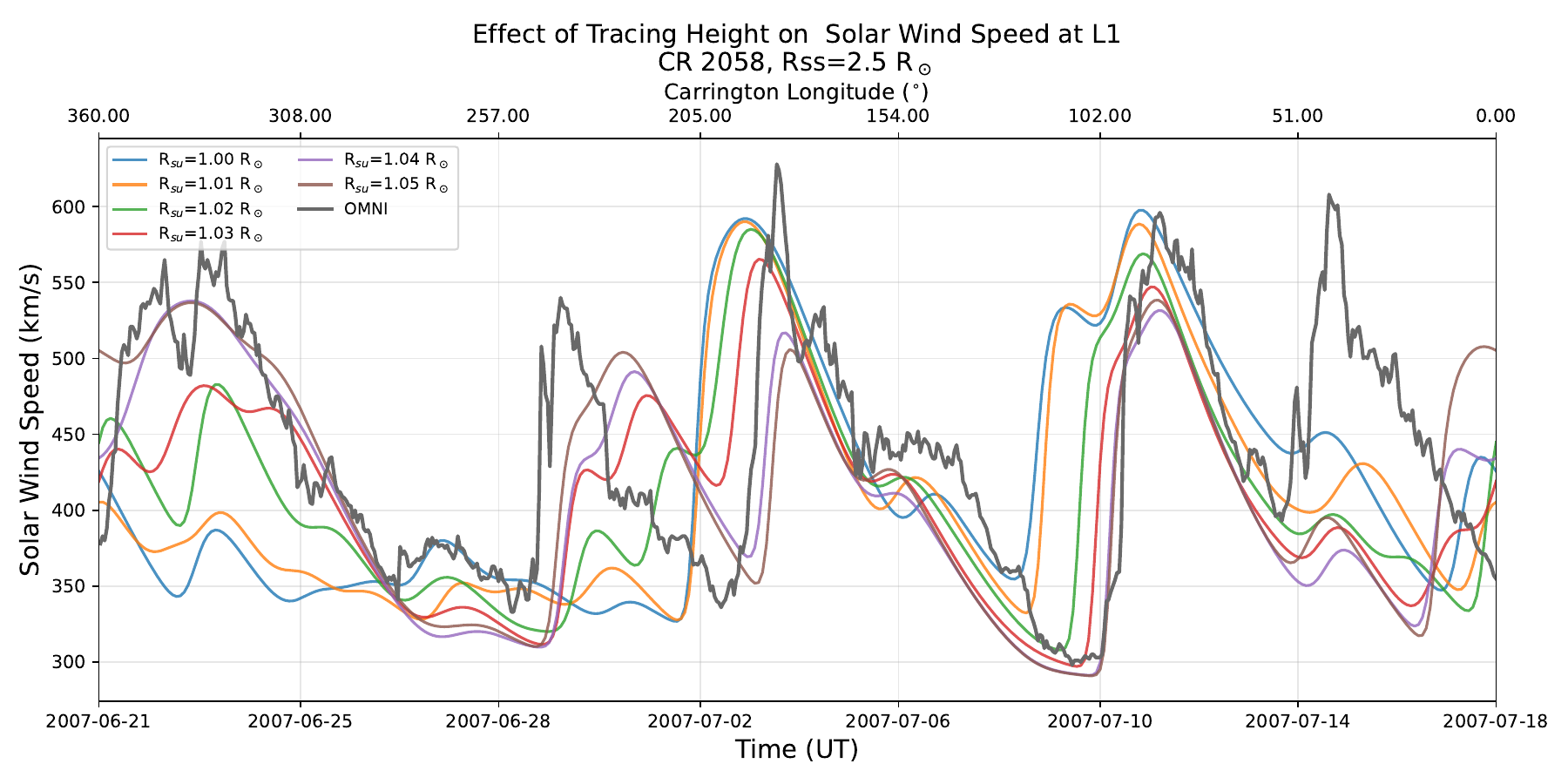}
\caption{Solar wind speed at L1 for CR 2058 for Rss = 2.5 R$_{\odot}$ and initial field-line tracing heights of 1.00 to 1.05 in steps of 0.01 shown in different colored lines. The grey line denotes the hourly averaged observed speed from the OMNI database.}
\label{cr2058_1}
\end{figure*}

\section{Results} \label{results}

Figure~\ref{cr2058_1} shows the observed solar wind speed profile for CR 2058 from OMNI (grey) together with a set of modeled solar wind speed profiles at L1 computed using a standard source surface height (Rss) of 2.5 R$_{\odot}$ and six different initial field-line tracing heights (R$_{su}$) ranging from 1.00 R$_{\odot}$ to 1.05 R$_{\odot}$ in steps of 0.01 R$_{\odot}$ (shown in colored lines). For the standard case of R$_{su}$ = 1.00 R$_{\odot}$, the model produces only two of the five observed HSEs. As R$_{su}$ is increased, we notice many improvements. First, the model begins to recover the first two HSEs that were missed under the standard configuration, and the predicted v$_{peak}$ values show a clear improvement with higher R$_{su}$ for those peaks. This indicates that increasing the tracing height can enhance the model’s ability to capture true HSEs that are otherwise undetected. Secondly, for the second, third and fourth HSEs, increasing R$_{su}$ leads to a noticeable reduction in the error of the predicted t$_{start}$, which in turn improves the estimation of t$_{dur}$. For CR 2958, raising the tracing height improves the model performance across all three key aspects of HSE prediction - t$_{start}$, t$_{dur}$, and v$_{peak}$, while simultaneously improving the prediction of missed HSE peaks.

\begin{figure*}[h]
\centering
\includegraphics[width=0.7\textwidth]{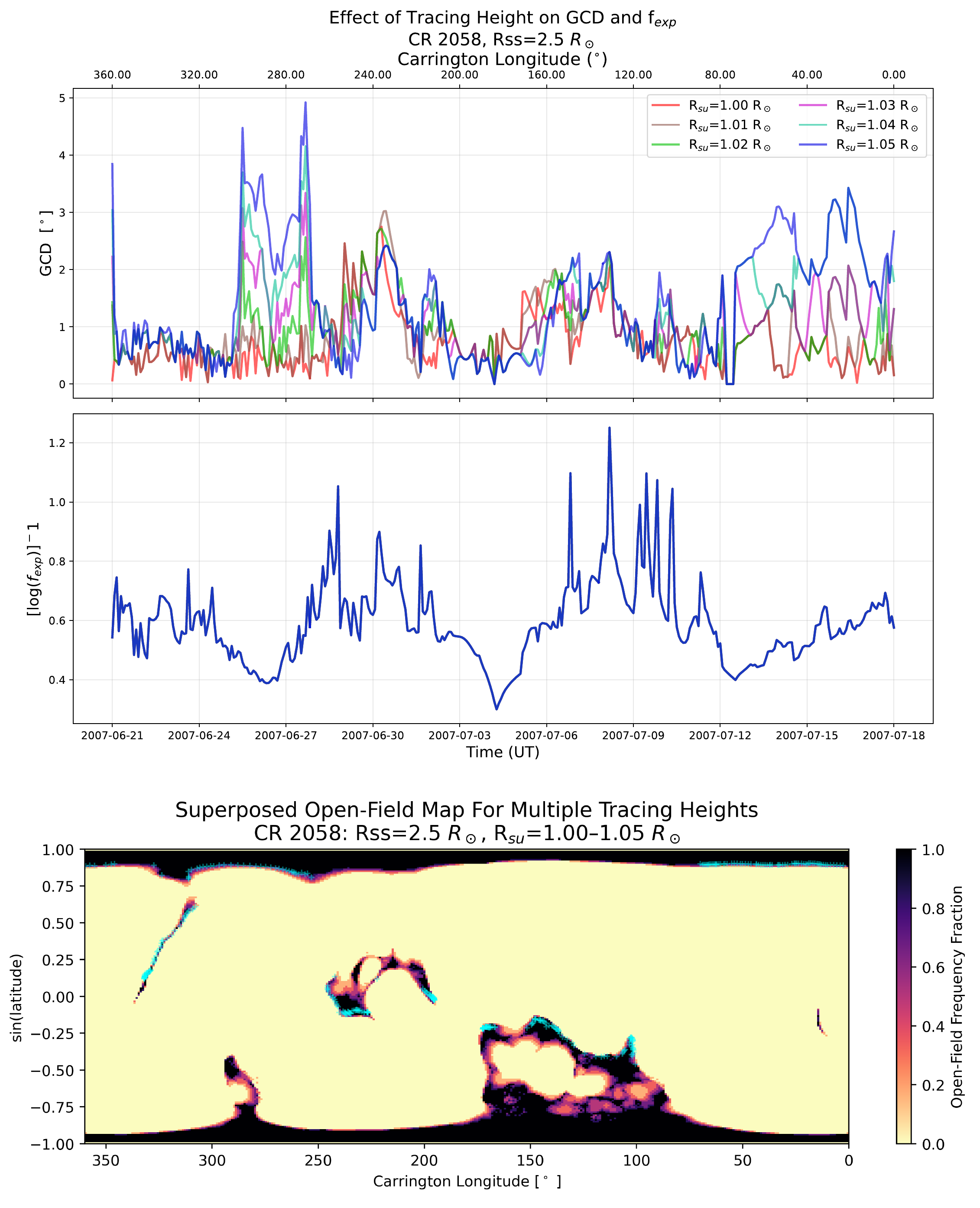}
          
\caption{(a) Effect of change in tracing height (for Rss = 2.5 R$_{\odot}$) for CR 2058, on the GCD profile (top panel) and inverse of the logarithm of the expansion factor profile (bottom panel) at sub-Earth locations at Rss. (b) An overlap of the modeled open field regions for 6 different tracing heights (for Rss = 2.5 R$_{\odot}$) for the same CR. The blue crosses indicate the footpoints of the field lines connected to the sub-Earth points.}
\label{cr2058_2}
\end{figure*}

To investigate the reasons for these improvements in CR 2058, we show in Figure~\ref{cr2058_2}(a), the GCD profiles at the sub-Earth locations for different values of R$_{su}$ (top panel), together with the corresponding $f_{exp}$ profiles (bottom panel) for the same CR. The GCD profiles exhibit substantial variation with changing R$_{su}$. In particular, higher R$_{su}$ values produce additional and more pronounced peaks, which directly correspond to the high-speed streams that were previously missed at L1. In contrast, the $f_{exp}$ profiles remain largely unchanged across the different R$_{su}$ values, indicating that the improvement in the modeled solar wind arises primarily from changes in GCD. To better understand why modifying R$_{su}$ leads to significant changes in the GCD distribution, Figure~\ref{cr2058_2}(b) presents a superposed map of the modeled open field regions for a fixed Rss of 2.5 R$_{\odot}$ and for the six different R$_{su}$ values. The colormap shows a open-field frequency fraction which essentially indicates the fractional occurrence of a pixel as magnetically open for different values of R$_{su}$. Thus, a value of 1 would indicate a region which was found to be magnetically open for all values of R$_{su}$. Figure~\ref{cr2058_2}(b) shows that the open-field boundaries are highly sensitive to the choice of R$_{su}$. As R$_{su}$ increases, the boundaries of the modeled open-field regions expand and become noticeably smoother. At the same time, small-scale fragmentation within the open-field regions, such as that seen in the large southern polar open-field region, is significantly reduced. The footpoints of field lines connected to the sub-Earth locations are marked with blue crosses. As R$_{su}$ is increased, these footpoints become associated with smoother and more extended open-field regions, which results in systematically larger and more stable GCD values. This shift in GCD ultimately translates into improved prediction of HSEs at L1, explaining the enhanced performance observed in the solar wind speed profiles.

\begin{figure*}[!htbp]
\centering
\includegraphics[width=0.98\textwidth]{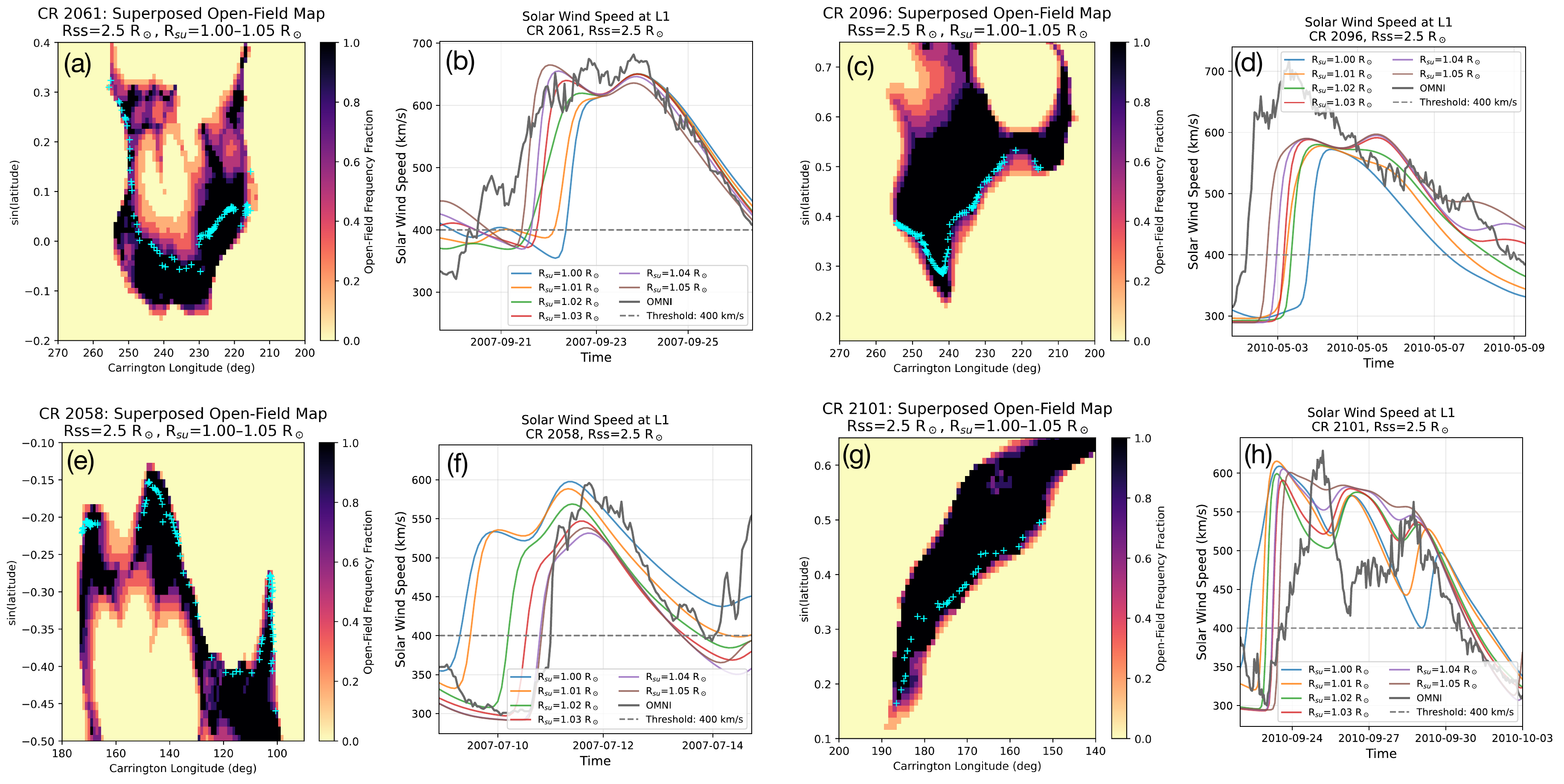}
\caption{(a,c,e,g) Cropped superposed modeled open-field regions for all R$_{su}$ values for Rss 2.5 R$_{\odot}$. Darker regions indicate higher overlap of open field regions for different cases of R$_{su}$. The blue crosses indicate the footpoints of the field lines connected to the sub-Earth points. (b,d,f,h) Corresponding HSEs that originate from these open-field regions. The grey profile is the observed speed from OMNI and the different colored lines indicate modeled speeds for different R$_{su}$ values. }
\label{t_dur}
\end{figure*}

\subsection{Effect of Field-line Tracing Height on t$_{start}$ and t$_{dur}$}

To further understand how increasing R$_{su}$ improves the prediction for t$_{start}$ and t$_{dur}$, Figure~\ref{t_dur}(a) shows, for CR 2061, a cropped superposed map of the modeled open-field region for different R$_{su}$ values, while panel (b) presents the corresponding HSE originating from this region. Similar examples of modeled open-field regions and their associated HSEs are shown for CR 2096 in panels (c) and (d), for CR 2058 in (e) and (f), and for CR 2101 in (g) and (h), respectively. In the open-field region maps, the footpoints of the field lines connected to the sub-Earth locations are marked with blue crosses. We find that with increasing R$_{su}$, the boundaries of the modeled open-field regions systematically expand, with clear extensions visible along both the eastern (left) and western (right) edges. This boundary extension, which is consistently observed across all cases, directly translates into corresponding shifts in the modeled starting and ending times of the associated HSEs, leading to an improved estimation of t$_{dur}$. Thus, as the tracing height is raised, the extension of the modeled open-field boundaries modulates the predicted onset and termination times of the HSEs, thereby reducing discrepancies in both t$_{start}$ and t$_{dur}$. From Figure~\ref{t_dur}, the maximum difference in the predicted t$_{start}$ across the tested R$_{su}$ values reaches approximately 1.5 days, which corresponds to nearly 50$\%$ of the total solar wind transit time from the Sun to L1, further indicating the importance of the R$_{su}$ parameter.

\begin{figure*}[!htbp]
\centering
\includegraphics[width=0.98\textwidth]{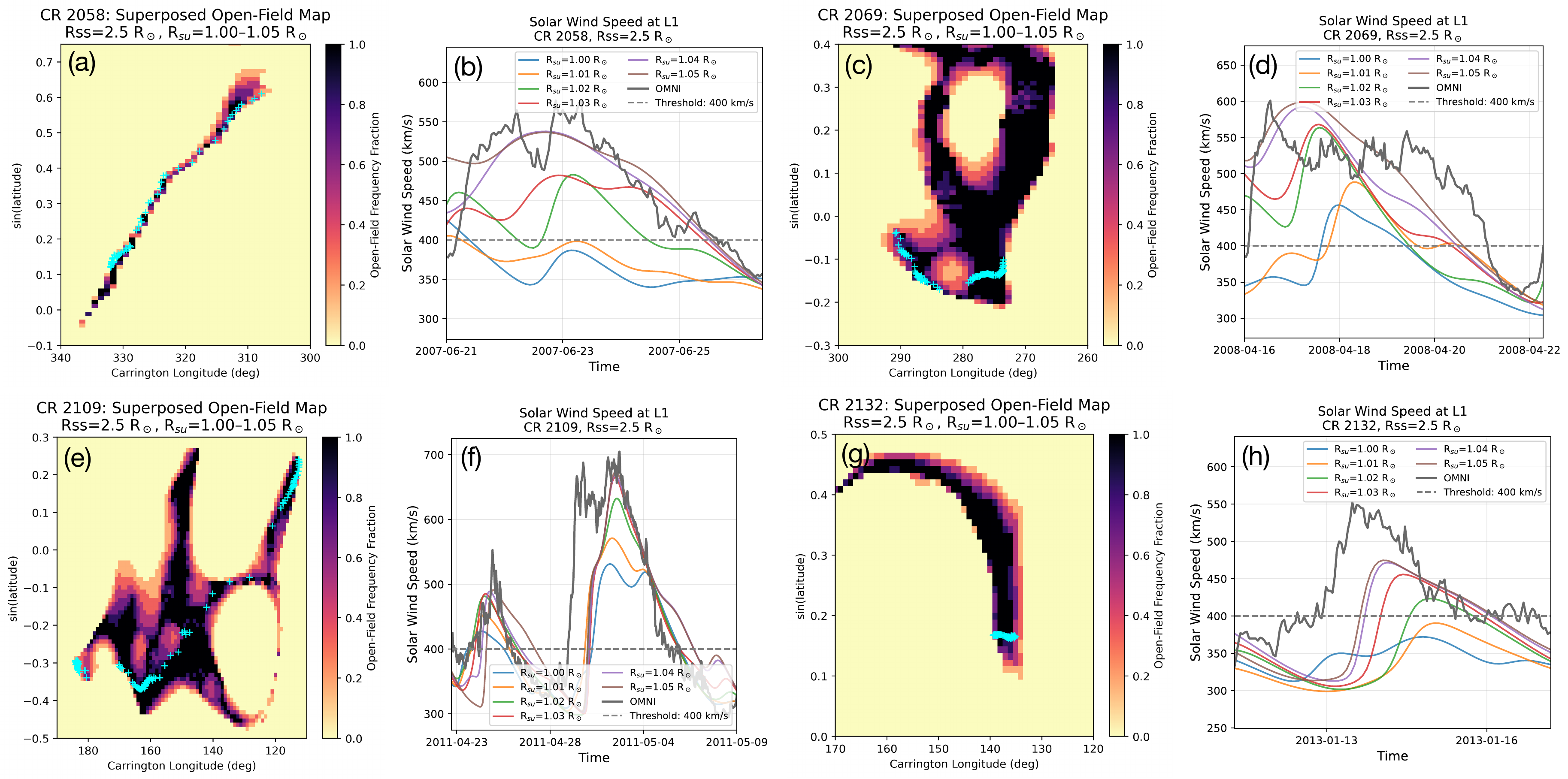}
\caption{Same as Figure~\ref{t_dur}.}
\label{true_pos}
\end{figure*}

\subsection{Effect of Field-line Tracing Height on Missed Peaks and v$_{peak}$}

In Figure~\ref{true_pos}, we again study the improvements in WSA predictions in the context of v$_{peak}$ and missed HSEs. We plot the cropped superposed maps of the modeled open-field regions and the corresponding modeled HSEs for different R$_{su}$ values for CR 2058 (in panels a, b), CR 2069 (panels c, d), CR 2109 (panels e, f) and CR 2132 (panels g, h) respectively. In these cases, increasing R$_{su}$ enables the model to predict HSEs that were completely missed when using the standard value of R$_{su}$ = 1.00 R$_{\odot}$ (see panels b, d, and h). For CR 2109 (panels e, f), the model already predicts the HSE at the standard tracing height, but with a large error in v$_{peak}$ of approximately 200 km s$^{-1}$; this discrepancy is substantially reduced as R$_{su}$ is increased. Inspection of the corresponding open-field region maps reveals that increasing R$_{su}$ produces progressively smoother and less fragmented open-field boundaries (e.g., panels a and g), particularly for small equatorial open-field regions and for regions exhibiting internal substructure within larger open-field regions (panels c and e). This reduction in boundary fragmentation suppresses spurious fine-scale features, leading to more coherent open-field geometries. As a result, the derived GCD values become more stable and less sensitive to small-scale morphological artifacts. These improvements directly reduce model uncertainty in magnetic footpoint mapping and, by extension, in the solar wind speed inferred through the WSA relation, especially in cases where fragmented or poorly defined open-field boundaries would otherwise introduce large variability in GCD. For example, in panel (a), we find that the open-field region is very thin and thus the GCD values for the footpoints associated with this open-field region would be very small. With a higher R$_{su}$, the boundary of this open-field region is relatively enhanced and that improves the GCD values associated with the footpoints, which can be seen as an improvement in the predicted speed at L1. Again, for CR 2069 and 2109, it can be seen that apart from the extension in the boundaries along the eastern and western edges (as discussed previously), there is also substantial internal fragmentation inside the open-field regions, especially near the footpoints of the field lines connected to the sub-Earth points. Thus, although the field lines remain rooted inside the open-field regions, this fragmentation reduces the GCD values. Increasing R$_{su}$ progressively suppresses these internal discontinuities. This leads to a much more stable calculation of the GCDs at the foot-point locations, which eventually gets reflected into much higher predicted speeds at L1, thereby improving the detection of missed peaks and reducing the error in v$_{peak}$. Thus, it seems that the connection between modeled open-field regions, modeled solar wind speed at L1 and a change in R$_{su}$ occurs in a twofold way. First, a higher R$_{su}$ extends the boundaries of the modeled open-field regions, which leads to improvement in t$_{start}$ and t$_{dur}$. Second, a higher R$_{su}$ reduces the inner fragmentation of open-field regions that leads to capturing missed HSEs and better estimation of v$_{peak}$. Further discussion on the relevance of these improvements in the context of the different components of Equation~\ref{eqn-v_wsa} are presented in Section~\ref{wsa_coeff_disc}.

\begin{figure*}[!htbp]
\centering
\includegraphics[width=0.9\textwidth]{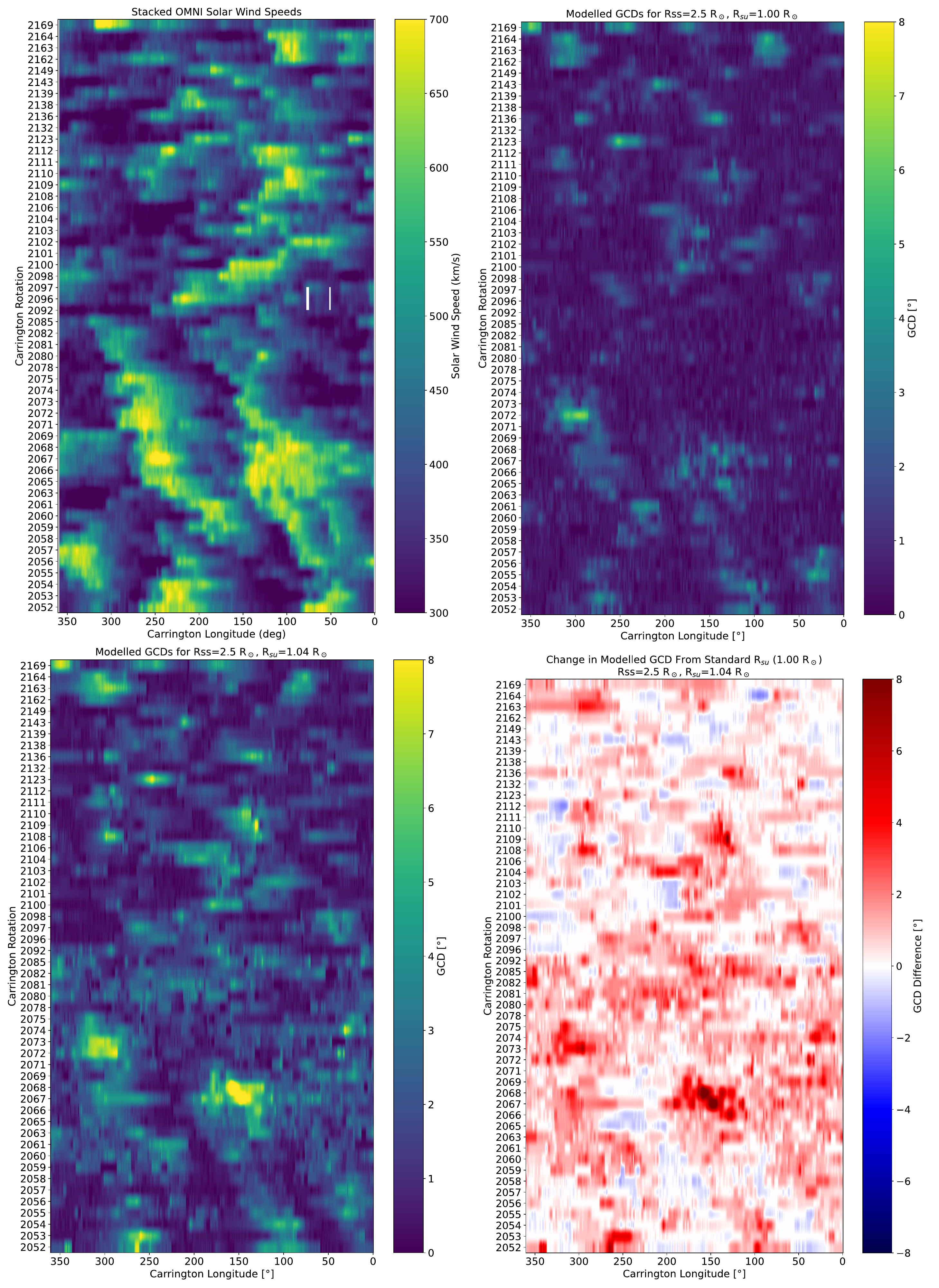}
\caption{Top left: Stacked plots of the observed solar wind speed at L1 from OMNI for 53 CRs between 2052 and 2169 are shown. Top right: Stacked plots of the computed GCDs at sub-Earth locations for the same CRs for a Rss = 2.5 R$_{\odot}$ and for R$_{su}$ of 1.0 R$_{\odot}$. Bottom left: Same as top right but for R$_{su}$ of 1.04 R$_{\odot}$. Bottom right: Change in GCD between R$_{su}$ of 1.00 and 1.04 R$_{\odot}$.}
\label{stack_CRs}
\end{figure*}

\subsection{Statistical Significance on Multiple Carrington Rotations}

To understand the effect of a higher R$_{su}$ on a larger sample of events, we repeated the analysis for 53 different CRs between CRs 2052-2169 (that corresponds to the time period between January 2007 - October 2015). CRs containing major ICMEs were excluded from the analysis using the updated ICME catalogue by \citeA{moestl2026ApJ}, as such transients strongly perturb the ambient solar wind structure and introduce non-stationary disturbances that are not captured by our modeling framework. In Figure~\ref{stack_CRs}, we plot stacked observed solar wind speed at L1 from OMNI for all 53 CRs (in the top left panel). In the top right and bottom left panels, we plot the modeled GCD profiles at the projected sub-Earth locations at Rss for R$_{su}$ of 1.00 R$_{\odot}$ and 1.04 R$_{\odot}$ respectively. The results show that a higher R$_{su}$ leads to a much better prediction of the peaks associated with the observed HSEs. We find that the magnitude of the peak GCD values at the locations of different HSEs have increased, indicating a higher estimate of v$_{peak}$. For better identification of the changes in GCD, the bottom right panel also shows the change in GCD between the R$_{su}$ 1.00 and 1.04 R$_{\odot}$ cases. For example, notable enhancements occur near Carrington longitudes of 150° during CR 2067–2069 and near 300° during CR 2072–2075. Thus, across a statistically significant set of rotations, increasing R$_{su}$ produces consistent modifications in GCD that translate into measurable improvements in the predicted solar wind profile at L1. The overall trend in the observed speeds is also better represented with a higher R$_{su}$ value, especially at the locations of the observed HSEs. The model-predicted solar wind speeds for all these CRs for different R$_{su}$ values are also made available (https://doi.org/10.6084/m9.figshare.33476032). It should be noted here that the analysis is based on a limited set of CRs, which may not fully capture the full variability of solar magnetic conditions across different phases of the solar cycle. Thus, it would be important to extend this analysis in future to a larger data set that would capture different phases of the solar cycle.

\begin{figure*}[!htbp]
\centering
\includegraphics[width=0.95\textwidth]{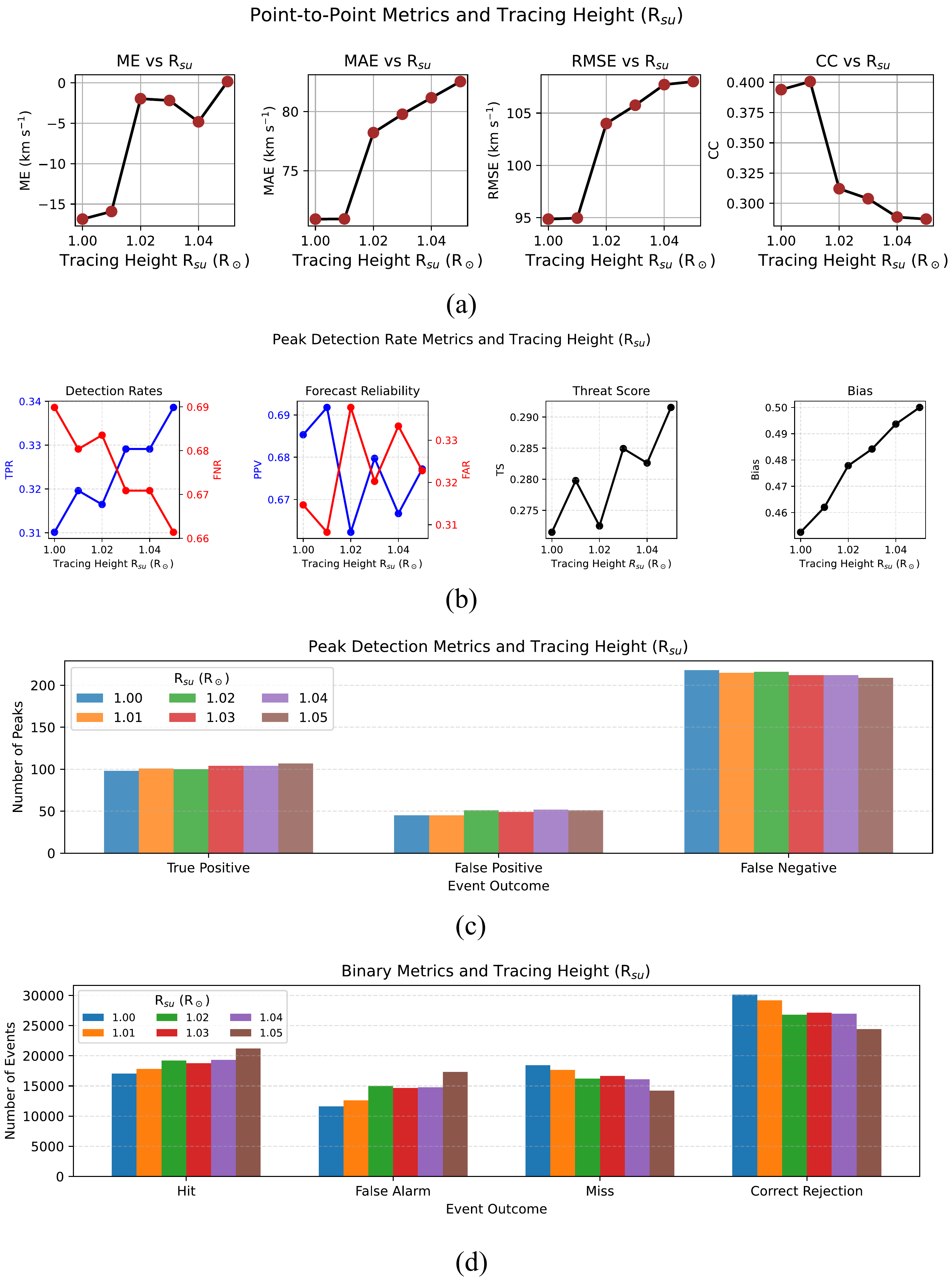}
\caption{Effect of change in R$_{su}$ for 53 CRs, with Rss = 2.5 R$_{\odot}$ on various point-to-point metrics (mean error, mean absolute error, root mean squared error and correlation coefficient) is shown in (a) and for various peak detection rate metrics (true positive and false negative rate, positive prediction value and false alarm ratio, threat score and bias) in (b). Statistical distribution of the number of hits, misses and false alarms in the peak detections for different values of R$_{su}$ is shown in (c) and the same for binary metrics (hits, misses, false alarms and correct rejections) is shown in for a threshold speed of 400 km/s (d).}
\label{metrics}
\end{figure*}

\begin{figure}
    \centering
    \includegraphics[width=0.9\linewidth]{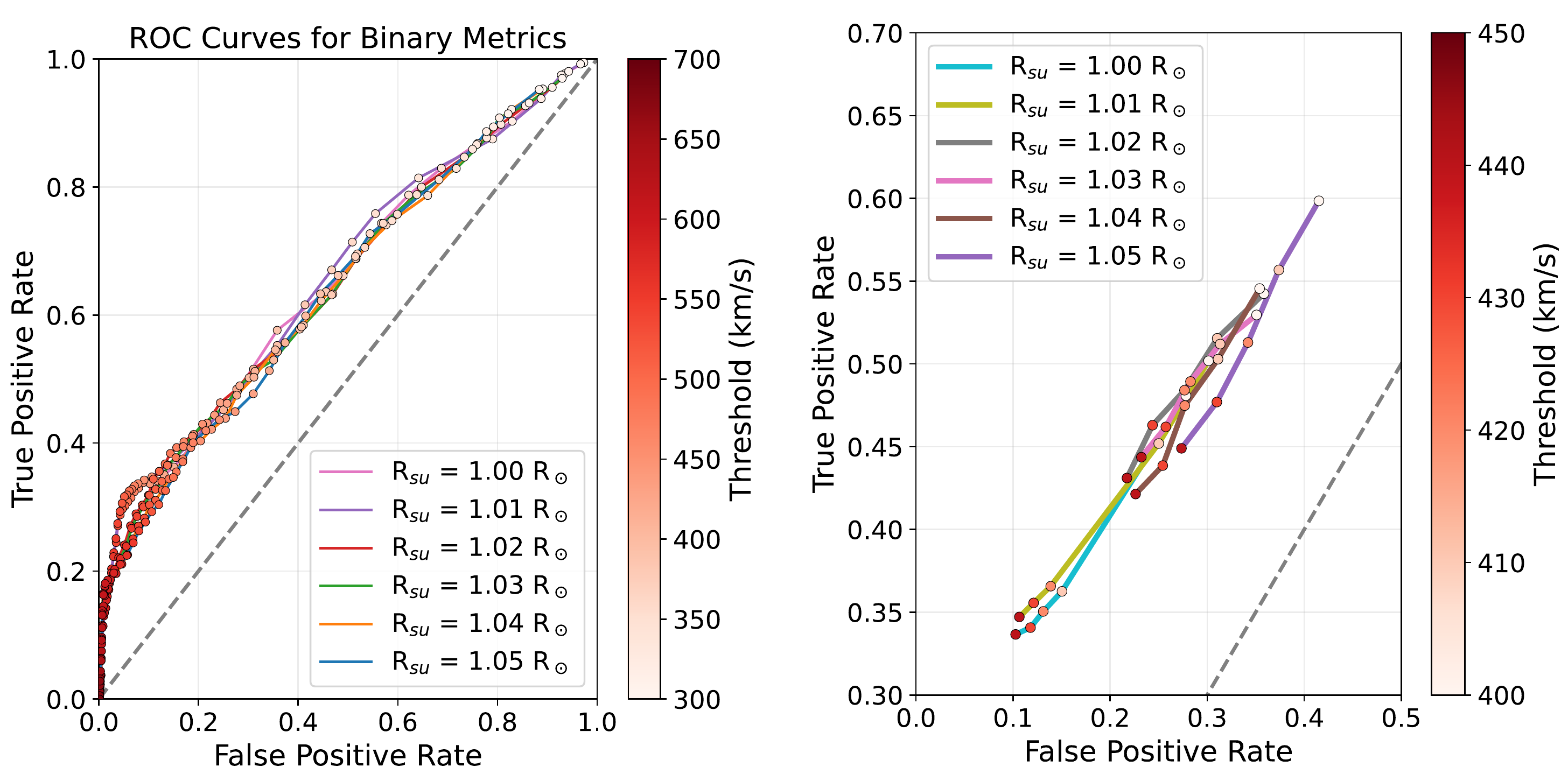}
    \caption{Left Panel: Receiver Operating Characteristic (ROC) curves showing the TPR versus the FPR for binary metrics for different R$_{su}$ values. The grey dashed line is for a random classifier. Right Panel: A plot of the TPR versus the FPR for a physically meaningful threshold range. }
    \label{roc}
\end{figure}

Now to quantify the improvement in solar wind predictions at L1, we employ a multi-metric approach with the entire dataset of 53 CRs. Firstly, in Figure~\ref{metrics}(a), we use different point-to-point metrics (that compare the observed and modeled values at each time step and provide an assessment of the overall agreement between the model and observations) like the Mean Error (ME), Mean Absolute Error (MAE), Root Mean Squared Error (RMSE) and the Correlation Coefficient (CC) and plot them with respect to R$_{su}$ for the whole dataset. We find that the ME decreases by around 15 km/s while there is a slight increase in the other errors ( 12 km/s in MAE, 11 km/s in RMSE) with increasing R$_{su}$ and a very small reduction in the CC. Thus, although we do not see any improvement in the point-to-point metrics (except for ME), the deterioration is not significant either. We next use peak detection metrics  \cite<as reported in>[]{Reiss2016SpWea}, which focus on the model's ability to capture solar wind speed peaks, and instead of evaluating every time step, this method identifies significant peaks in the modeled and observed speeds inside a given time window and compares them. We define some key peak detection terms as True Positive (TP): when the model correctly predicts an observed peak, False Positive (FP): when the model predicts a peak for no corresponding observed peak and False Negative (FN): when the model fails to predict an observed peak. Using these quantities, we define the various peak detection metrics such as the True Positive Rate (TPR) and the False Negative Rate (FNR) as

\begin{equation}
    TPR\,=\, \frac{TP}{TP+FN};\quad FNR\,=\,\frac{FN}{TP+FN}
\end{equation}

We define a Positive Prediction Value (PPV) which measures the fraction of the total predicted peaks that were present in observations and a False Alarm Ratio (FAR) which indicates the fraction of the total predicted peaks that were not present in observations as follows:

\begin{equation}
    PPV\,=\, \frac{TP}{TP+FP};\quad FAR\,=\, \frac{FP}{TP+FP}
\end{equation}

We also define a Threat Score (TS) which evaluates the overall detection skill while penalizing for both missed and false peaks and a Bias which quantifies whether the model tends to over-predict (Bias $>$ 1) or under-predict (Bias $<$ 1) the number of peaks as follows:

\begin{equation}
    TS\,=\, \frac{TP}{TP+FP+FN};\quad Bias\,=\, \frac{TP+FP}{TP+FN}
\end{equation}

Thus, in Figure~\ref{metrics}(b), we plot these metrics versus R$_{su}$ for all 53 CRs. We find that the detection rates improve with higher R$_{su}$, as the TPR increases and FNR decreases with higher R$_{su}$. In terms of testing the reliability of the prediction, we find that the PPV and the FAR does not show any steady change but fluctuates. We do find improvement in TS and the Bias indicates that although the model is under-predicting, the under-prediction is steadily reduced with higher R$_{su}$. This improvement in peak detections is further represented in numbers (instead of fractions) in Figure~\ref{metrics}(c), where we plot bar charts of the TPs, FPs and FNs and it can again be seen that the model predicts more TPs with higher R$_{su}$ while the number of FPs fluctuates with a general trend of increasing. The FNs again decrease with higher R$_{su}$, thus showing improvement in detecting more peaks and minimising the number of missed peaks. Thus, instead of comparing each point in time of the predicted solar wind profile to the observed profile, the peak detection metrics help in assessing the ability of a model in predicting the HSEs that were observed in a given time window.

We also apply a set of binary event metrics which evaluate how well models predict the occurrence or non-occurrence (in a yes or no output) of events defined by a threshold value. By comparing every point in time, this approach classifies outcomes into categories such as Hits - event both predicted and observed, Miss - event observed but not predicted,  False Alarm - event predicted but not observed  and Correct Rejection - no event predicted and none observed. We first plot the Receiver Operating Characteristic (ROC) curves in Figure~\ref{roc} left panel, that neatly summarizes the predictive capabilities by plotting the TPR versus the FPR for specific thresholds. The grey dashed line indicates the same for a random classifier that has equal TPR and FPR. We note that all R$_{su}$ values perform better than a random classifier, but there is no clear signature of the best R$_{su}$ value for the entire threshold range. We thus again plot the TPR versus the FPR for a physically meaningful threshold range in Figure~\ref{roc} right panel and we find that a threshold of 400 km/s produces the best outcome and a R$_{su}$ of 1.05 R$_{\odot}$ produces the best outcome for that threshold. We thus consider a threshold of 400 km/s and in Figure~\ref{metrics}(d), we again plot a similar distribution of these four quantities. We see an increase in the number of hits for higher values of R$_{su}$ compared to the standard value, with the highest number of hits for the highest R$_{su}$. On the other hand, we see a general increase in false alarm, a general decreasing trend in the number of misses and correct rejections.  Overall, for binary metrics, we see  clear and steady trends of improvement in the number of hits and misses, however a higher R$_{su}$ seems to do worse in terms of false alarms and correct rejections. This also indicates the importance of how sensitive the point-to-point comparison metrics are to a temporal shift in the predicted profiles, with respect to the observed profile, and the usefulness of peak-detection-based metrics in terms of quantifying the improvement in HSE prediction capabilities. Now, although increasing R$_{su}$ improves predictive performance, an optimal choice of R$_{su}$ may be case-dependent and would require further investigation across a broader range of datasets that would encompass different phases of the solar cycle. Apart from the aspects discussed above, it should be noted that this study does not aim to investigate the physical origins of solar wind formation, but rather focuses on improving empirical forecasting performance within the WSA modeling framework

\begin{figure*}[!htbp]
\centering
\includegraphics[width=0.9\textwidth]{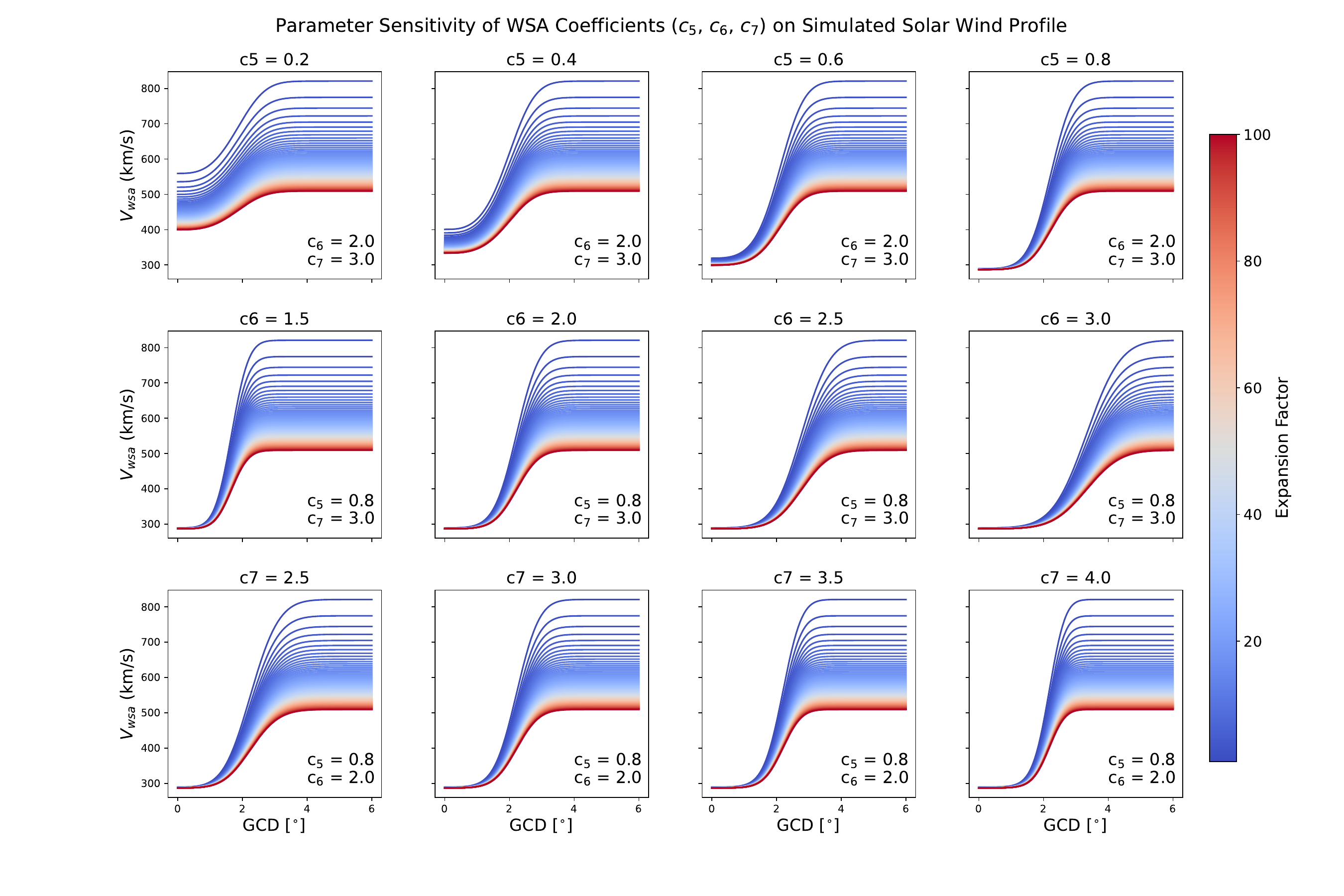}
\caption{Effect of change in different WSA coefficients for a simulated input data of $f_{exp}$ and GCD. We vary the $f_{exp}$ between 1 and 100 (indicated by color coded lines) and GCD between 0$^{\circ}$-$\,6^{\circ}$. The effect of varying the coefficients c$_5$, c$_6$, and c$_7$ are shown in the top, middle and bottom rows respectively.}
\label{wsa_param}
\end{figure*}

\section{Discussion} \label{discussions}

\subsection{Relevance to WSA Coefficient Tuning:}\label{wsa_coeff_disc}
Having seen the different aspects of improvement obtained by raising the R$_{su}$, it is important to place these results in the context of previous efforts that have sought to enhance WSA predictions through optimization of its empirical coefficients \cite{McGregor2008JGRA,Riley2015SpWea,Reiss2020ApJ,Mayank2025ApJ}. In our analysis, we find that varying R$_{su}$ primarily modifies the GCD profile, while the EFs remain largely unchanged (Figure~\ref{cr2058_2}). This naturally motivates a focused investigation of the DCHB component of Equation~\ref{eqn-v_wsa}, with the aim of comparing the effects of coefficient variations within this term to those arising from changes in R$_{su}$. As a first step, it is necessary to isolate and understand the role of the key coefficients governing the GCD dependence. To this end, we perform a targeted parameter sensitivity study of the coefficients c$_5$, c$_6$ and c$_7$ using synthetic inputs spanning $f_{exp}$ values from 1 to 100 and GCD values between 0$^{\circ}$ and 6$^{\circ}$, consistent with the range observed in our dataset. A similar analysis on the coefficients c$_6$ and c$_7$ was performed in \citeA{Mayank2022ApJS} and \citeA{Samara2024ApJ}, but we also include c$_5$ in our study for a more complete comparison. Figure~\ref{wsa_param} summaries the effect of the three coefficients on the predicted WSA speed profile.

Focusing on the DCHB component of Equation~\ref{eqn-v_wsa}, it is evident that c$_5$ primarily influences the small-d regime, i.e., regions close to the open-field boundary. For larger d, the exponential term rapidly approaches zero, and v$_{wsa}$ transitions to the asymptotic high-speed regime where the influence of c$_5$ (as well as c$_6$ and c$_7$) becomes negligible and it is mainly the $f_{exp}$ that makes the difference. Thus, c$_5$ effectively controls the strength of the slow wind suppression near the boundary; that is, larger values of c$_5$ lead to stronger suppression and consequently slower winds near the boundary. While the $f_{exp}$ also modulates the speed at small d, its influence becomes progressively less pronounced as c$_5$ increases, as illustrated in the top row of Figure~\ref{wsa_param}. In contrast, c$_6$ acts as a characteristic length scale that determines the angular extent of the transition from slow to fast wind. Increasing c$_6$ broadens this transition region, delaying the approach to the asymptotic fast wind regime and producing a more gradual variation in speed with d, as seen in Figure~\ref{wsa_param}. Physically, at the open-field boundary, this corresponds to an expansion of the open/closed boundary region, such that fast wind speeds are reached only deeper within the open-field interior. From a forecasting perspective, this implies that c$_6$ plays a key role in controlling the predicted arrival time of high-speed streams. Finally, c$_7$ governs the sharpness of the transition within the window set by c$_6$. Larger values of c$_7$ result in a steeper, more abrupt transition from slow to fast wind, whereas smaller values produce a smoother, more gradual change. This behavior is clearly reflected in Figure~\ref{wsa_param}, where increasing c$_7$ leads to a more pronounced “knee” in the speed profile, while the width of the transition remains unchanged.

In the context of the present study, increasing R$_{su}$ leads to improved estimates of t$_{start}$ and t$_{dur}$ as well as a better prediction of the slow wind regime, indicating that it partially reproduces the effects associated with both c$_6$ and c$_7$. In addition to this, two further improvements are observed: an enhanced detection of previously missed high-speed streams (along with improved estimates of v$_{peak}$), and a better characterization of the slow wind regime, reflected in higher correct-rejection values (Figure~\ref{metrics}). These improvements can be understood in terms of how R$_{su}$ modifies the GCD distribution. In cases of highly fragmented modeled open-field boundaries, a significant fraction in the open/closed field map is assigned very small d values. As discussed earlier, this is precisely the regime where the influence of c$_5$ is strongest, leading to enhanced suppression of the wind speed. As a result, regions that should contribute to fast wind can get incorrectly mapped to slow wind, causing missed HSSs and reduced peak speeds. Increasing R$_{su}$ mitigates this effect by smoothing and spatially extending the open-field boundaries, thereby reducing the occurrence of small d values. This, in turn, decreases the extent to which the c$_5$-driven suppression is activated, allowing the model to recover fast wind streams that were previously damped. In this sense, a higher R$_{su}$ effectively reduces the local effect of c$_5$, not by altering its value, but by shifting the input distribution away from the regime in which its influence dominates. Thus, in a nutshell, some of the improvements introduced by a higher R$_{su}$ are qualitatively similar to those achieved through tuning the coefficients c$_5$, c$_6$ and c$_7$ in WSA solar wind predictions at L1. It is also worth noting that the DCHB term in Equation~\ref{eqn-v_wsa} is non-linear in d, implying that small and abrupt variations in d, particularly near open-field boundaries, can lead to large changes in v$_{wsa}$. Thus, ensuring a stable and spatially coherent distribution of d would lead to more stable estimation of v$_{wsa}$ and hence more stable solar wind predictions at Earth. By simplifying the large-scale open/closed boundary, a higher R$_{su}$ essentially makes the model more robust by reducing the extent to which the model operates within highly complex open-field boundaries.


\subsection{Relevance to Coronal Model Optimization:}

The results of this study also have direct relevance to the recently explored path of source surface height (Rss) optimization in PFSS-based solar wind modeling. Varying Rss is known to significantly impact the large-scale coronal magnetic field structure, leading to changes in both the $f_{exp}$ and GCD profiles \cite{Majumdar2025ApJ}, and thereby influencing the predicted solar wind speed. However, reducing Rss below the standard value of 2.5 R$_{\odot}$ often produces modeled open-field boundaries that significantly exceed the observed boundaries, and it often introduces extra artificial open-field regions that are not supported by observations \cite{Lee2011SoPh,Asvestari2019JGRA,Nikolic2019SpWea}. On the other hand, raising the source surface to much higher values than 2.5 R$_{\odot}$ could also lead to much smaller modeled open-field boundaries than the observed boundaries and could also lead to missing smaller open-field regions. Thus, it seems that while adjusting the source surface can improve agreement with in-situ measurements at L1, they do so at the cost of compromising the actual physical realism in the modeled coronal structure. One possible approach to mitigating this issue is to relax the assumption of a spherical source surface and instead consider a non-spherical geometry \cite{Levine1982SoPh,Schulz1997AnGeo,Kruse2020A&A}.

On the contrary, this study shows another alternative strategy of keeping the source surface at its standard height and instead varying R$_{su}$, which allows the GCD distribution to be modified without substantially altering the global coronal topology. This approach thus avoids the additional complexities associated with changing the source surface height and, at the same time, produces more coherent modeled open-field boundaries, leading to improved predictions of solar wind speed at L1. As demonstrated in this study, increasing R$_{su}$ primarily acts to smooth and extend the boundaries of existing open-field regions, while largely preserving their overall morphology and, most importantly, without introducing any extra artificial open-field regions. This distinction is critical, as it allows for improvements in solar wind prediction—such as better HSS timing, enhanced peak speeds, and reduced missed events—without invoking significant unphysical changes to the large-scale coronal magnetic field. By retaining Rss at its standard value and instead varying R$_{su}$, the present approach also offers a more balanced pathway to improving predictive skill while preserving physical consistency, thereby helping the trade-offs commonly associated with Rss-based optimization. Furthermore, \citeA{Majumdar2025ApJ} showed that changes in the model grid resolution can modify the modeled boundaries of open-field regions, which in turn can affect the prediction accuracy of HSEs. However, simply increasing or decreasing the model resolution does not smooth out the internal fragmentation associated with small-scale closed-field regions embedded within larger open-field regions (see Figure 2 in \citeA{Majumdar2025ApJ}). The effect of increasing \(R_{\rm su}\) is therefore qualitatively distinct from that of changing the model grid resolution.

In agreement with our findings, it is also interesting to note that in a previous study, \citeA{Jones2024ApJ} showed that using a slightly higher tracing height of 1.01 R$_{\odot}$ first allows the model to get rid of the ``parasitic closed-flux regions" that can arise from the spherical harmonic (SH) approximation used in the Arge et al. code for constructing the coronal magnetic field from photospheric observations. Secondly, they reported that using 1.01 R$_{\odot}$ as the tracing height yields improved consistency in both modeled solar wind speeds and open field regions between the SH-based approach and the PFSS solver POT3D, which employs a finite-difference method. In our study, we also use \textit{pfsspy} which uses a finite difference solver to model the coronal field. Thus, this further suggests that R$_{su}$ can serve as a physically motivated and complementary parameter that reduces numerical artifacts, leads to better agreement between different coronal field reconstruction methods (i.e. spherical harmonics and finite difference methods) and thereby improves WSA solar wind predictions at L1.
 
\subsection{Importance of Multi-Metric Evaluation of Model Performance:}

In a space weather forecasting context, the primary objective is not only to reproduce the observed solar wind speed time series, but also to accurately predict the arrival, duration, and amplitude of HSEs. Evaluating model performance solely through commonly used point-to-point metrics such as ME, MAE, RMSE, and correlation coefficient provides a measure of overall agreement between modeled and observed values, but does not fully capture these event-based aspects of prediction skill \cite{Owens2005JGRA}. In particular, such metrics can obscure (or even unfairly penalize the model for) discrepancies in the timing and amplitude of HSEs, as they treat all time steps equally and do not distinguish between background solar wind conditions and HSEs. To address these limitations, in line with the recent recommendations in \citeA{Reiss2023AdSpR} from the H1-01 Ambient Solar Wind validation Team embedded in the  Committee on Space Research (COSPAR) - International Space Weather Action Teams (ISWAT) initiative, we adopted a multi-metric evaluation that complements traditional point-to-point diagnostics with peak detection and binary classification metrics. The peak detection approach \cite{Reiss2016SpWea}, comparable to previous algorithms reported in \citeA{Owens2005JGRA} and \citeA{MacNeice2009SpWea-II,MacNeice2009SpWea-I}, specifically assesses the model’s ability to identify and reproduce significant speed enhancements beyond a threshold value, within defined time windows, thereby directly evaluating its performance in capturing HSE structures. In parallel, we also used binary metrics that quantify the model’s skill in predicting the occurrence or non-occurrence of an event based on a threshold criterion for every time step. 

In this regard, we have noted in this study that by raising R$_{su}$, the model often creates more pronounced peaks, or produces peaks that were previously missed with the standard R$_{su}$. Here, it should be noted that the recovered missed peaks were produced from the same open-field regions from the standard R$_{su}$ (where a higher R$_{su}$ produced smoother and spatially extended large-scale open-field boundaries) and not by introducing new open-field regions as sources of recovered peaks. These peaks produced with a higher R$_{su}$ could have a timing offset relative to the observed t$_{start}$ and it seems like the model pays a price for that in terms of the point-to-point metrics. However, from a space weather perspective, the model is actually doing a better job in terms of predicting missed peaks or improving the predictions in terms of timing or amplitude of the HSEs, and this is reflected in the other event-based metrics or the binary metrics. Nevertheless, it should be noted that such event detection algorithms also have their own limitations. For example, in peak-detection metrics, the evaluation depends on the choice of threshold speed and the temporal window used for comparison, while in binary metrics the outcome is again strongly controlled by the selected threshold level. In the case of binary metrics, timing offsets remain a challenge, but there is a distinction from point-to-point metrics. A predicted HSE that is shifted in t$_{start}$ relative to observations could lead to a significant error in point-to-point metrics, but for binary metrics, it would still be counted as a “hit”, provided the prediction is beyond the threshold value. Thus, although, the timing offset challenge remains for binary metrics yet the penalty associated with such timing offsets is considerably reduced compared to point-to-point evaluations. As a result, event-based metrics provide a complementary and often more operationally relevant perspective on model performance. This also highlights an important aspect for space-weather forecasting: it is generally more valuable for a model to predict the occurrence of an HSE with some offset in t$_{start}$ than to miss the event entirely, even if the former may yield larger errors in point-to-point metrics.  However, keeping these limitations of different metrics in mind, it is thus equally important to examine the full time series of modeled and observed solar wind speeds to contextualize the metric-based evaluation. Thus, for completeness, time series corresponding to all CRs analyzed in this study are made available (https://doi.org/10.6084/m9.figshare.33476032) for reference. Together with visual inspection of the time series, the use of a multi-metric framework is particularly important in the context of the present study, as the improvements obtained by increasing R$_{su}$ manifest differently across various aspects of model performance, which would not be highlighted by the use of any one specific choice of metrics. 

\subsection{Towards Real-Time Implementation}
In the context of real-time space weather prediction, we plan to further test the proposed R$_{su}$-based optimization approach using a setup mimicking that currently running in the real-time WSA framework implemented in the Community Coordinated Modeling Center (CCMC) WSA dashboard (https://ccmc.gsfc.nasa.gov/wsa-dashboard/) to assess its performance. It should be noted that the configuration running at the CCMC differs in certain aspects from the framework used in this study. For example, in the standard real-time framework, the PFSS model is coupled with a Schatten Current Sheet model \cite{Schatten1969SoPh} which is then used in conjunction with Equation~\ref{eqn-v_wsa} to get the near-Sun speed map, that is finally propagated outward using a one dimensional solar wind model. In this transition from historic CR-based analyses to real-time prediction, several other practical considerations need to be taken into account, including data latency and availability, as well as uncertainties associated with frequently updated input magnetic maps in evolving solar conditions. Such a transition will allow us to assess the robustness and predictive skill of this proposed R$_{su}$-based optimization in a true real-time setting.


\section{Summary and Conclusion} \label{conclusions}

In this study, we investigated a new physically motivated approach to improving the predictive performance of the widely used WSA model, with a particular focus on forecasting HSEs at L1. Our modeling framework combines a PFSS model to represent the large-scale coronal magnetic field, the WSA empirical formulation to derive near-Sun solar wind speeds from the reconstructed coronal magnetic field, and the HUX model to propagate these near-Sun speeds to L1. The presence of highly fragmented structures in modeled open-field maps often complicates the identification and representation of large-scale boundaries that are critical for accurate WSA predictions. Such fragmentation often arises from small-scale closed field regions embedded within otherwise open field regions, and these structures are typically confined to low coronal heights. Motivated by this, we examined the effect of raising the initial field line tracing height (R$_{su}$) above the photosphere as a means to reduce the influence of these small-scale features. Specifically, instead of adopting the standard value of 1.00 R$_{\odot}$ as the inner boundary for field line tracing, we considered six values of R$_{su}$ in the range 1.00–1.05 R$_{\odot}$, in increments of 0.01 R$_{\odot}$, thereby effectively tracing field lines above the heights where such small-scale closed structures are most prevalent. This allows us to assess whether reducing open-field region fragmentation in this manner leads to measurable improvements in solar wind prediction accuracy at L1. The main conclusions of this study are summarized below:

\begin{enumerate}
    \item Raising R$_{su}$ produces significant changes in the predicted solar wind speed at L1, leading to improved accuracy in predicting HSEs (Figure~\ref{cr2058_1}). These improvements arise primarily from modifications to the GCD distribution, while the $f_{exp}$ remains largely unchanged (Figure~\ref{cr2058_2}, top and middle panels). Analysis of the superposed modeled open-field regions further shows that increasing R$_{su}$ systematically modifies the large-scale open-field boundaries, making them smoother and more spatially extended. We also find that a R$_{su}$ of 1.05 R$_{\odot}$ effectively reduces the internal fragmentation caused by small-scale closed-field regions within the larger open-field regions, suggesting \(1.05\,R_{\odot}\) as a reasonable upper limit for R$_{su}$.

    \item Improvements in HSE prediction accuracy are observed across multiple aspects. In particular, a higher R$_{su}$ leads to better estimates of t$_{start}$ and t$_{dur}$ of HSEs, which are two key characteristics of these events. These improvements are primarily attributed to the extension of large-scale open-field boundaries (resulting from a higher R$_{su}$) from which the HSEs originate (Figure~\ref{t_dur}).

    \item Further improvements are observed in v$_{peak}$ of HSEs, as well as in the detection of previously missed peaks. This enhancement is particularly evident in cases where the open-field regions exhibit highly complex morphologies with significant boundary and internal fragmentation. In such situations, small-scale closed loop structures lead to large numbers of pixels in the open-field map being assigned artificially small GCD values. Increasing R$_{su}$ mitigates this effect by producing more spatially coherent and extended open-field boundaries, resulting in more stable GCD distributions and ultimately enabling the recovery of previously suppressed peak speeds.

    \item Applying our methodology to a set of 53 different CRs, we found similar enhancements in GCD profiles during the occurrence of HSEs, thus indicating the overall applicability of this method. The predictive performance of the model is further assessed using multiple evaluation metrics. While standard point-to-point metrics do not show any clear improvement with increasing R$_{su}$ (except for ME), event-based metrics such as peak detection and binary classification metrics provided a better characterization of model skill in terms of HSE predictions. In particular, higher R$_{su}$ leads to improved prediction rates of HSEs, and reduced false alarms,  thereby demonstrating enhanced event-based forecasting performance. 

    \item A higher R$_{su}$ also helps capture the combined roles played by the coefficients c$_5$, c$_6$ and c$_7$ in Equation~\ref{eqn-v_wsa}. Moreover, unlike Rss optimization, which significantly modifies the global coronal magnetic field and can introduce artificial open-field regions, variations in R$_{su}$ primarily affect the boundary representation of existing open-field regions without adding extra spurious open-field regions. Thus, using a higher R$_{su}$ offers a useful alternative to other existing WSA optimization efforts. 

    \item We believe that the interpretation of a higher R$_{su}$ would partly depend on the scientific context and objective of the study. In the context of the present work, we primarily interpret a higher R$_{su}$ as a practical modeling approach for mitigating sensitivities associated with small-scale structures within the modeled open-field regions and, consequently, for obtaining a more robust determination of the large-scale open/closed-field boundaries. This interpretation is particularly relevant in cases where such small-scale structures do not themselves play a significant role in determining the solar wind prediction accuracy. Whether the use of a higher R$_{su}$ also corresponds to a physically more appropriate height to extract information from the PFSS solution is a separate question. Addressing this possibility would require further investigation and supporting evidence.
\end{enumerate}

This study thus presents a strategy that, has not been systematically investigated previously for improving HSE prediction accuracy within a widely used PFSS-based solar wind modeling framework. It also addresses recent recommendations outlined in the new COSPAR Space Weather Roadmap to advance our ambient solar wind prediction capabilities \cite{KUZNETSOVA2026}. We expect that the results presented here may contribute toward more accurate solar wind predictions at L1. It should be noted that this study demonstrates the improvement in solar wind prediction accuracies at Earth using higher R$_{su}$ values, while the recommendation for an optimal R$_{su}$ value would need further investigation on a much larger dataset. For future studies, it will be important to investigate whether the optimal R$_{su}$, in terms of solar wind prediction accuracy, exhibits any systematic dependence on solar-cycle phase. Such an analysis, based on a sufficiently extensive dataset spanning different phases of the solar cycle, would further help determine whether a universal optimal value of R$_{su}$ can be identified or whether the optimal tracing height varies over the solar cycle. As a next step, we plan to implement and test this approach in real-time, using a framework that is similar to the one running in the WSA dashboard hosted by the CCMC. This would enable us to evaluate the robustness and predictive skill of this approach under real-time conditions, beyond historic CRs.

\section{Open Research}
The photospheric magnetic field data used in this study were obtained from the publicly available Global Oscillation Network Group (GONG) archive at https://gong.nso.edu/. Hourly averaged OMNI data was used from the OMNI database which is publicly available at https://omniweb.gsfc.nasa.gov/ow.html. The coronal magnetic field extrapolations were performed using the open-source Python package \textit{pfsspy} \cite{Stansby2020}, while the heliospheric propagation of the solar wind was carried out using the openly available Heliospheric Upwind eXtrapolation (HUX) code \cite{Riley2011SoPh}. For the analysis of the data, different python packages were used which are openly available. These packages include numpy \cite{harris2020numpy}, pandas \cite{reback2020pandas}, astropy \cite{astropy2022}, matplotlib \cite{hunter2007matplotlib} and sunpy \cite{Mumford2022zndo}. The model-predicted solar wind speeds for all these CRs for different R$_{su}$ values are also made available (https://doi.org/10.6084/m9.figshare.33476032). 


\section*{Conflict of Interest Statement}
The authors have no conflicts of interest to disclose.

\acknowledgments
The authors thank the anonymous reviewers for the
comments that have helped improve the manuscript. This work uses GONG data obtained by the NSO Integrated Synoptic Program, managed by the National Solar Observatory, which is operated by the Association of Universities for Research in Astronomy (AURA), Inc. under a cooperative agreement with the National Science Foundation and with contribution from the National Oceanic and Atmospheric Administration. The GONG network of instruments is hosted by the Big Bear Solar Observatory, High Altitude Observatory, Learmonth Solar Observatory, Udaipur Solar Observatory, Instituto de Astrofísica de Canarias, and Cerro Tololo Inter-American Observatory. This research was funded in whole or in part by the Austrian Science Fund (FWF) [10.55776/P34437]. For open access purposes, the author has applied a CC BY public copyright license to any author-accepted manuscript version arising from this submission. S.M. acknowledges Yi-Ming Wang for his valuable comments on this study.


%
\bibliography{references_sw} 
%




%
%
%
%
%

\end{document}